\documentclass[12pt]{article}
\pdfoutput=1

\usepackage{scicite}

\usepackage{times}

\usepackage{graphicx}

\newenvironment{sciabstract}{%
\begin{quote} \bf}
{\end{quote}}

\newcounter{lastnote}

\usepackage{authblk}
\usepackage{xcolor}
\usepackage[colorlinks=true,linkcolor=blue,urlcolor=blue,citecolor=blue,bookmarksopenlevel=2,bookmarksopen=true]{hyperref}
\usepackage[labelfont=bf]{caption}
\usepackage{amsmath}
\usepackage{amssymb}
\usepackage{aas_macros_long}
\usepackage{booktabs}
\usepackage{lscape}
\usepackage{longtable}
\usepackage{threeparttablex}
\usepackage{xspace}
\newcommand{\msun}{{\rm M}_{\odot}}

\newcommand{\tdor}{\mbox{30~Dor}\xspace}

\newcommand{\farr}{Farr and Mandel\@\xspace}

\newcommand{\mytitle}{Response to comment on ``An excess of massive stars in the local 30~Doradus starburst''}
\title{\mytitle} 

\author[1$\ast$]{F.R.N. Schneider}
\author[2]{H. Sana}
\author[3]{C.J. Evans}
\author[4,5]{J.M. Bestenlehner}
\author[6]{N. Castro}
\author[7]{L. Fossati}
\author[8]{G. Gr{\"a}fener}
\author[8]{N. Langer}
\author[3]{O.H. Ram{\'i}rez-Agudelo}
\author[9]{C. Sab{\'i}n-Sanjuli{\'a}n}
\author[10,11]{S. Sim{\'o}n-D{\'i}az}
\author[12]{F. Tramper}
\author[5]{P.A. Crowther}
\author[13,2]{A. de Koter}
\author[13]{S.E. de Mink}
\author[14]{P.L. Dufton}
\author[15]{M. Garcia}
\author[16]{M. Gieles}
\author[17,18]{V. H\'{e}nault-Brunet}
\author[10,11]{A. Herrero}
\author[19,16]{R.G. Izzard}
\author[20]{V. Kalari}
\author[12]{D.J. Lennon}
\author[21]{J. Ma\'{i}z Apell\'{a}niz}
\author[22]{N. Markova}
\author[15]{F. Najarro}
\author[1,8]{Ph. Podsiadlowski}
\author[23]{J. Puls}
\author[3]{W.D. Taylor}
\author[24]{J.Th. van Loon}
\author[25]{J.S. Vink}
\author[26,27]{C. Norman}

\affil[1]{\normalsize{Department of Physics, University of Oxford, Keble Rd, Oxford OX1 3RH, United Kingdom}}
\affil[2]{\normalsize{Institute of Astrophysics, KU Leuven, Celestijnenlaan 200D, 3001, Leuven, Belgium}}
\affil[3]{\normalsize{UK Astronomy Technology Centre, Royal Observatory Edinburgh, Blackford Hill, Edinburgh EH9 3HJ, United Kingdom}}
\affil[4]{\normalsize{Max-Planck-Institut f{\"u}r Astronomie, K{\"o}nigstuhl 17, 69117 Heidelberg, Germany}}
\affil[5]{\normalsize{Department of Physics and Astronomy, Hicks Building, Hounsfield Road, University of Sheffield, Sheffield S3 7RH, United Kingdom}}
\affil[6]{\normalsize{Department of Astronomy, University of Michigan, 1085 S. University Avenue, Ann Arbor, MI 48109-1107, USA}}
\affil[7]{\normalsize{Austrian Academy of Sciences, Space Research Institute, Schmiedlstra{\ss}e 6, 8042 Graz, Austria}}
\affil[8]{\normalsize{Argelander-Institut f{\"u}r Astronomie der Universit{\"a}t Bonn, Auf dem H{\"u}gel~71, 53121~Bonn, Germany}}
\affil[9]{\normalsize{Departamento de F{\'i}sica y Astronom{\'i}a, Universidad de La Serena, Avda. Juan Cisternas $N^o$ 1200 Norte, La Serena, Chile}}
\affil[10]{\normalsize{Instituto de Astrof{\'i}sica de Canarias, E-38205 La Laguna, Tenerife, Spain}}
\affil[11]{\normalsize{Departamento de Astrof{\'i}sica, Universidad de La Laguna, E-38206 La Laguna, Tenerife, Spain}}
\affil[12]{\normalsize{European Space Astronomy Centre, Mission Operations Division, PO Box 78, 28691 Villanueva de la Ca\~nada, Madrid, Spain}}
\affil[13]{\normalsize{Astronomical Institute Anton Pannekoek, Amsterdam University, Science Park 904, 1098 XH Amsterdam, The Netherlands}}
\affil[14]{\normalsize{Astrophysics Research Centre, School of Mathematics and Physics, Queen's University Belfast, Belfast BT7 1NN, Northern Ireland, United Kingdom}}
\affil[15]{\normalsize{Centro de Astrobiolog\'ia (CSIC-INTA), Ctra. de Torrej\'on a Ajalvir km-4, E-28850 Torrej\'on de Ardoz, Madrid, Spain}}
\affil[16]{\normalsize{Department of Physics, Faculty of Engineering and Physical Sciences, University of Surrey, Guildford, GU2 7XH, United Kingdom}}
\affil[17]{\normalsize{National Research Council, Herzberg Astronomy \& Astrophysics, 5071 West Saanich Road, Victoria, BC, V9E 2E7, Canada}}
\affil[18]{\normalsize{Department of Astrophysics/IMAPP, Radboud University, PO Box 9010, NL-6500 GL Nijmegen, The Netherlands}}
\affil[19]{\normalsize{Institute of Astronomy, The Observatories, Madingley Road, Cambridge CB3 0HA, United Kingdom}}
\affil[20]{\normalsize{Departamento de Astronom{\'i}a, Universidad de Chile, Camino El Observatorio 1515, Las Condes, Santiago, Casilla 36-D, Chile}}
\affil[21]{\normalsize{Centro de Astrobiolog{\'i}a, CSIC-INTA, ESAC campus, camino bajo del castillo s/n, E-28\,692 Villanueva de la Ca\~nada, Spain}}
\affil[22]{\normalsize{Institute of Astronomy with National Astronomical Observatory, Bulgarian Academy of Sciences, PO Box 136, 4700 Smoljan, Bulgaria}}
\affil[23]{\normalsize{Ludwig-Maximilians-Universit{\"a}t M{\"u}nchen, Universit{\"a}tssternwarte, Scheinerstrasse 1, 81679 M{\"u}nchen, Germany}}
\affil[24]{\normalsize{Lennard-Jones Laboratories, Keele University, Staffordshire, ST5 5BG, United Kingdom}}
\affil[25]{\normalsize{Armagh Observatory, College Hill, Armagh, BT61 9DG, Northern Ireland, United Kingdom}}
\affil[26]{\normalsize{Johns Hopkins University, Homewood Campus, Baltimore, MD 21218, USA}}
\affil[27]{\normalsize{Space Telescope Science Institute, 3700 San Martin Drive, Baltimore, MD 21218, USA}\vspace{0.5cm}}

\affil[$\ast$]{\normalsize{To whom correspondence should be addressed; E-mail: fabian.schneider@physics.ox.ac.uk.}}

\date{}

\begin{document}


\baselineskip24pt


\maketitle 



\begin{sciabstract}
\farr reanalyse our data, finding initial-mass-function slopes for high mass stars in 30~Doradus that agree with our results. However, their reanalysis appears to underpredict the observed number of massive stars. Their technique results in more precise slopes than in our work, strengthening our conclusion that there is an excess of massive stars above $30\,\msun$ in 30~Doradus.
\end{sciabstract}

\farr \cite{Farr} reanalysed the results of our study \cite{2018Sci...359...69S}, in which we investigated the star-formation history (SFH) and stellar initial mass function (IMF) of the local 30~Doradus (\tdor) starburst in the Large Magellanic Cloud and found an overabundance of stars with initial masses beyond $30\,\msun$. They use an alternative and potentially more powerful statistical framework, hierarchical Bayesian inference, and infer IMF power-law indices for massive stars that are in agreement with our results (compare the IMF slope distributions in their fig.~1 to the $1\sigma$ range inferred in our analysis). Their analysis allows them to infer the IMF slope with higher precision than was possible in our case, such that their inferred IMF slope for high mass stars in \tdor is shallower than that of a Salpeter IMF \cite{1955ApJ...121..161S} with an even larger confidence (more than 95.5\% compared to 83\% in our analysis). Their reanalysis therefore supports our main findings and conclusions about the IMF in \tdor.

Farr's and Mandel's \cite{Farr} main criticism of our work is that ``[t]here is no statistical meaning to [age and mass] distribution[s] obtained'' by adding the posterior probability distributions of the ages and initial masses inferred for individual stars. It is true that such distributions are not posterior probability functions in a Bayesian framework. However, we caution that the IMF is historically defined as a histogram of stellar masses \cite{1955ApJ...121..161S,1979ApJS...41..513M,1986FCPh...11....1S,1993MNRAS.262..545K,2001MNRAS.322..231K,2003PASP..115..763C,2010ARA&A..48..339B} and our procedure to add the posterior probability distributions of the initial masses of individual stars is the equivalent of computing a histogram for the mass distribution of a sample of stars, while taking into account the observational uncertainties of individual mass estimates. Virtually all IMFs inferred in the literature are constructed in this way, so Farr's and Mandel's criticism implicitly applies to those as well. The VLT-FLAMES Tarantula Survey (VFTS) \cite{2011A&A...530A.108E} has reached a completeness of about 73\% with respect to a more complete census \cite{2013A&A...558A.134D} of massive stars in \tdor (see fig.~S2 in our original work). For a complete stellar sample, the age distribution of stars obtained with our method would directly provide the SFH at the youngest ages where even the most massive stars did not yet end their nuclear burning lifetime---so there is also meaning to age distributions constructed as was done in our work.

\begin{figure}
\centering
\includegraphics[width=0.65\textwidth]{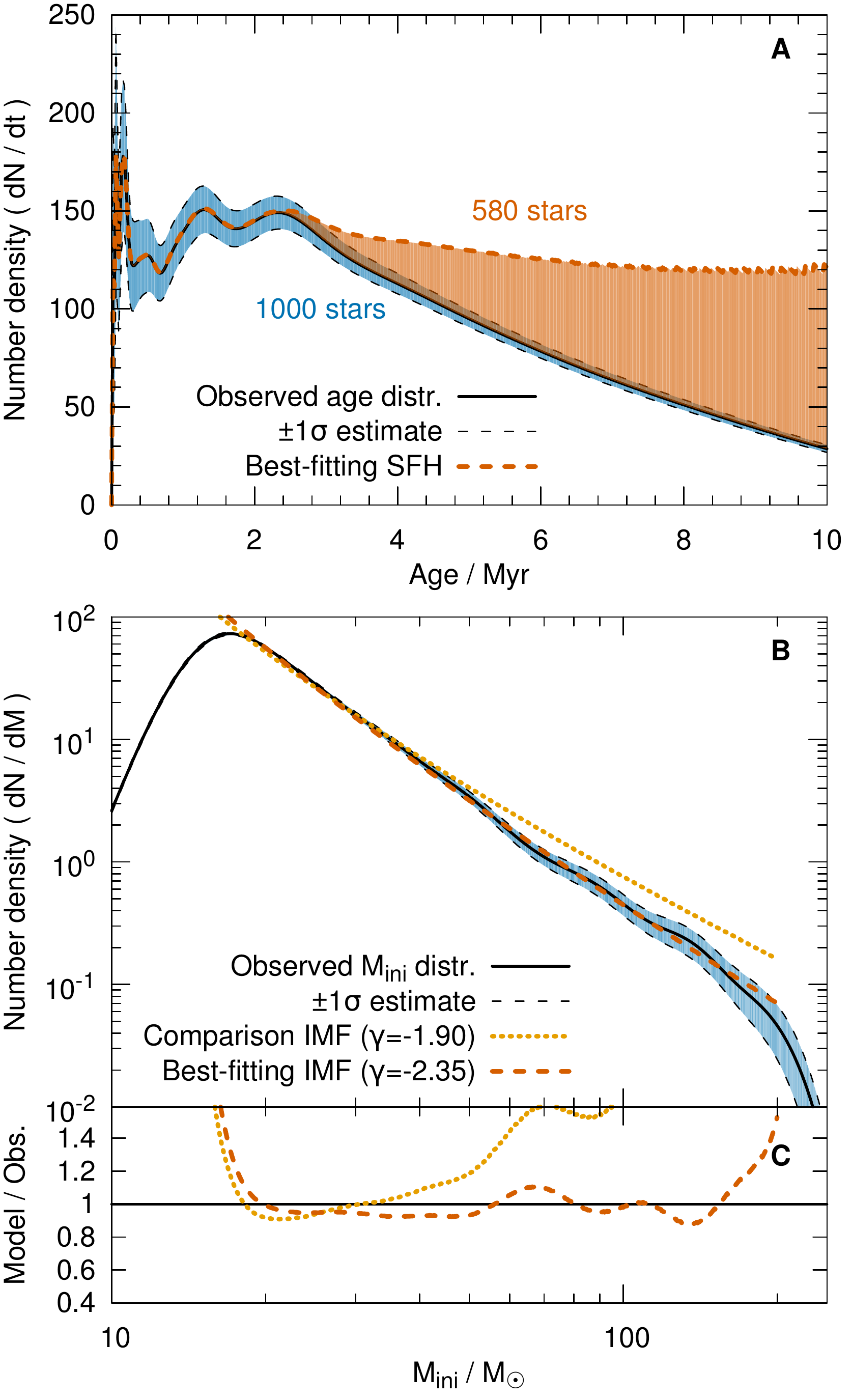}
\caption{\textbf{Inference of the SFH and IMF of a mock stellar population.}
Distributions of ages (A) and initial masses (B) of the mock stars (black lines) sampled from a Salpeter IMF with slope $\gamma=-2.35$ including bootstrapped $1\sigma$ estimates. The best-fitting IMF and SFH are indicated by the red dashed lines. For comparison, the predicted distribution of initial masses is shown for an IMF slope of $\gamma=-1.90$ (orange dotted line). C) Ratio of the predicted model and ``observed'' mock initial mass distributions, showing that the two IMF models only deviate from the mock data by more than the uncertainty above $30\,\msun$.}
\label{fig:sfh-imf-mock-data}
\end{figure}

We have tested our statistical analysis with mock data. To this end, we sampled a stellar population of 1000 stars more massive than $15\,\msun$ for a given Salpeter high-mass IMF with slope $\gamma=-2.35$ and a continuous SFH (constant star formation rate). In this way, we have obtained Gaussian distributions of the ages and masses of individual mock stars with $1\sigma$ uncertainties of 20\% and 15\% in age and mass, respectively. These uncertainties are characteristic of the age and mass uncertainties of stars in our sample of \tdor stars \cite{2018Sci...359...69S}. We then used exactly the same analysis technique as in our original work to infer the IMF and SFH of the mock star sample. The results of this test are shown in Fig.~\ref{fig:sfh-imf-mock-data} and demonstrate that our analysis method is able to reproduce the underlying SFH and IMF of the mock stars. For comparison, we show the distribution of initial masses for an IMF with slope $\gamma=-1.90$ to illustrate that our analysis technique can distinguish between a Salpeter IMF slope of $\gamma=-2.35$ and a shallower slope of $\gamma=-1.90$. This test further shows that both IMFs reproduce the mock data similarly well in the mass range $15\text{--}30\,\msun$ and that the high mass end (${>}\,30\,\msun$) of the distribution of mock masses has the largest power to constrain the high-mass IMF slope (Fig.~\ref{fig:sfh-imf-mock-data}C).

Our analysis of the VFTS data relies on two different techniques to infer the high-mass end of the IMF: (i) by fitting the observed distribution of stars in the mass range $15\text{--}200\,\msun$ and (ii) by fitting the number of stars more massive than $30$ and $60\,\msun$. Both procedures give results that are in good agreement \cite{2018Sci...359...69S}. From the inferred masses and corresponding uncertainties of our sample stars, we find $75.9^{+6.8}_{-7.0}$ stars above $30\,\msun$ and $22.2^{+4.0}_{-4.6}$ stars above $60\,\msun$ \cite{2018Sci...359...69S}. Contrarily to what \farr write in their reanalysis \cite{Farr}, their online data (\href{https://github.com/farr/30DorIMF}{https://github.com/farr/30DorIMF}, as accessed on 6th May 1pm GMT) suggest that their best-fitting SFH and IMF models underpredict the observed number of massive stars. They predict on average ${\approx}\,65$ stars above $30\,\msun$ and ${\approx}\,18$ stars above $60\,\msun$. Their ratio of the number of stars ${>}\,30\,\msun$ to the number of stars ${>}\,60\,\msun$ (${\approx}\,3.6$) is larger than what we have observed in \tdor (${\approx}\,3.4$), which appears to be consistent with \farr inferring slightly steeper IMF slopes than we did in our analysis. Indeed, using our SFH model and the results of our fitting method (ii), the numbers of massive stars above $30$ and $60\,\msun$ as predicted by \farr are found for an IMF slope of about $\gamma=-2.10$ (fig.~2 in our original work \cite{2018Sci...359...69S}). This is consistent with their best-fitting IMF slopes of $\gamma=-2.05$ to $-2.15$ for the different SFH models.

The reanalysis of \farr gives systematically steeper IMF slopes than in our work and consequently seems to underpredict the observed number of massive stars in \tdor. We do not know the cause of this discrepancy. Our methodology appears to be robust and the only other obvious difference in the two approaches---besides the statistical framework---is the assumption on the SFH. We directly infer the SFH from the data without making assumptions on its functional form. \farr assume Gaussian and exponential SFH models that provide more degrees of freedom than in our case, and find IMF slope differences of $\Delta \gamma \approx 0.1$ depending on the assumed SFH model. This is a systematic uncertainty that we did not discuss in our original work and that makes the inference of the IMF of composite stellar populations even more challenging.

\renewcommand{\refname}{References and Notes}
\bibliographystyle{../Science}


\paragraph*{Acknowledgements}

\textbf{Funding:} This work was supported by the Oxford Hintze Centre for Astrophysical Surveys which is funded through generous support from the Hintze Family Charitable Foundation. HS acknowledges support from the FWO-Odysseus program under project G0F8H6N. GG acknowledges financial support from the Deutsche Forschungsgemeinschaft, Grant No.\ GR 1717/5. OHRA acknowledges funding from the European Union's Horizon 2020 research and innovation programme under the Marie Sk{\l}odowska-Curie grant agreement No 665593 awarded to the Science and Technology Facilities Council. CS-S acknowledges support from CONICYT-Chile through the FONDECYT Postdoctoral Project No.~3170778. SSD and AH thank the Spanish MINECO for grants AYA2015-68012-C2-1 and SEV2015-0548. SdM has received funding under the European Union’s Horizon 2020 research and innovation programme from the European Commission under the Marie Sk{\l}odowska-Curie (Grant Agreement No. 661502) and the European Research Council (ERC, Grant agreement No. 715063). MGa and FN acknowledge Spanish MINECO grants FIS2012-39162-C06-01 and ESP2015-65597-C4-1-R. MGi acknowledges financial support from the Royal Society (University Research Fellowship) and the European Research Council (ERC StG-335936, CLUSTERS). RGI thanks the STFC for funding his Rutherford fellowship under grant ST/L003910/1. VK acknowledges funding from the FONDECYT-Chile fellowship grant No.\ 3160117. JMA acknowledges support from the Spanish Government Ministerio de Econom{\'i}a y Competitividad (MINECO) through grant AYA2016-75 931-C2-2-P. NM acknowledges the financial support of the Bulgarian NSF under grant DN08/1/13.12.2016. STScI is operated by AURA, Inc. under NASA contract NAS5-26555.

\textbf{Author contributions:} FRNS wrote the manuscript and all authors contributed to its discussion.

\textbf{Competing interests:} None


\clearpage

\end{document}